\documentclass[10pt,times new roman, onecolumn, twoside, letterpaper, journal, final]{IEEEtran}
\usepackage[pdftex]{graphicx}
\usepackage[cmex10]{amsmath}
\usepackage{array}
\usepackage{mdwmath}
\usepackage{multirow}
\usepackage{hyperref}

\usepackage{amsmath}
\usepackage[hang]{footmisc}
\usepackage{amsfonts}
\usepackage{amsmath}
\usepackage{array}
\usepackage{amssymb}
\usepackage{amsmath}
\usepackage{amssymb}
\usepackage{subdepth}
\usepackage{amsthm}
\usepackage{footnote}

\usepackage[super]{nth}
\usepackage{longtable}
\usepackage{cite}
\usepackage{color}

\usepackage{epstopdf}
\usepackage{caption}
\usepackage{subcaption}
  \usepackage{flushend}
  \usepackage[utf8]{inputenc}
\usepackage[english]{babel}

\newcommand{\tk}{\theta_k}
\newcommand{\PoutD}{P_\text{delay}^{\text{out}}}
\newcommand{\Dmax}{D^k_{\text{max}}}
\newcommand{\ECk}{E^k_{\text{c}}}

\begin{document}

\title{Flexible Multiple Access Enabling Low-Latency Communications: Introducing NOMA-R}

\author{ Mouktar Bello, Wenjuan Yu, Mylene Pischella, Arsenia Chorti, Inbar Fijalkow, Leila Musavian}
\maketitle

\section{Introduction}

Various verticals in 5G and beyond (B5G) networks require very stringent latency guarantees, while at the same time envisioning massive connectivity. As a result, choosing the optimal multiple access (MA) technique to achieve low latency is a key enabler of B5G. In particular, this issue is more acute in uplink transmissions due to the potentially high number of collisions. On this premise, in the present contribution we discuss the issue of  delay-sensitive uplink connectivity using optimized MA techniques; to this end, we perform a comparative analysis of various MA approaches with respect to the achievable effective capacity (EC). As opposed to standard \emph{rate} (PHY) or \emph{throughput} (MAC) analyses, we propose the concept of the effective capacity as a suitable metric for characterizing jointly PHY-MAC layer delays.

The palette of investigated MA approaches includes standard orthogonal MA (OMA) and power domain non orthogonal MA (NOMA) in uplink scenarios, both considering random pairing and optimized pairing alternatives. It further extends to encompass a recently proposed third alternative, referred to as NOMA-Relevant (NOMA-R), which extends OMA and NOMA approaches by flexibly selecting the MA technique. We show that optimizing both user pairing and MA selection increases the network EC, especially when stringent delay constraints are in place; thus a flexible MA is a potentially preferable strategy for future low latency applications.

\section{Low Latency B5G Landscape}
Future communication networks will not only support an evolution of traditional communication services, but also address novel verticals where real-time response is critical, e.g., autonomous vehicles or industry 4.0   \cite{FengVuceticURLLCMag,PopovskiIEEENetwork18}. While for these types of services, 5G and beyond (B5G) networks are targeting very stringent latency guarantees, e.g., between 1 and 10 ms, the network is required to provide other key performance indicators (KPIs) as well, for example, to provide high throughput in the context of ultra high speed low latency communications (uHSLLC), or to provide ultra reliability in the context of ultra reliable low latency communications (URLLC) \cite{PopovskiIEEENetwork18}. Meeting these conflicting requirements is a key to open up new business opportunities and benefit a number of novel use cases including remote surgery, self-driving vehicles and tactile Internet.

Recently, rich research interest has concentrated on these topics, as enabling true real-time connectivity seems to  be among the tallest hurdles in B5G fulfilment. So far, major efforts in addressing low latency requirements have been channeled to the empowerment of edge and fog computing on one hand, and, network slicing on the other, while in parallel novel radio access and scheduling approaches at the MAC sub-layer are actively studied, under the umbrella of flexible numerology \cite{FengVuceticURLLCMag}.

\subsection{Which delay metric to use?}

\par In this context, an interesting question arises with respect to characterizing delays jointly at the radio (PHY) and MAC layers. In fact, in B5G, quality of service (QoS) according to the class based system of the differentiated services (DiffServ) paradigm will be in place everywhere -- even for low-end Internet of things (IoT) sensors. In particular, in the first { \href{https://www.oulu.fi/6gflagship/news/6g-white-paper}{\textcolor{blue}{\underline{6G white paper}}}} it is noted that: `` \textit{6G needs an upgraded networking paradigm moving from best effort to differentiated service quality}''. At the same time, considerable amounts of traffic will be placed on the wireless edge (e.g., ``\textit{M2M connections will be more than half of the global connected devices and connections by 2022}'' as projected in the \href{https://www.cisco.com/c/en/us/solutions/collateral/service-provider/visual-networking-index-vni/white-paper-c11-741490.html}{\textcolor{blue}{\underline{Cisco Visual Networking Index}}: Forecast and Trends, 2017–2022 White Paper}.

In the late nineties, delays were explicitly characterised in asynchronous transfer mode (ATM) wired networks that employed integrated services (IntServ) QoS  by using the concept of the effective bandwidth \cite{Tse99} (later
extending to cover DiffServ). In the wireless MAC however, due to small scale fading and shadowing, it is inherently impossible to provide hard delay guarantees. Additionally, DiffServ QoS, by design, can only  provide statistical delay guarantees due to the absence of resource reservations with the exception of the network edges.

In wireless networks, to provide \textit{statistical} delay guarantees the concept of the  \textit{effective capacity (EC)} \cite{wu2003effective}, can be employed. It has been recently proven that a delay
bound and a maximal tolerable delay bound violation
probability can only be jointly satisfied if the effective capacity exceeds the effective bandwidth \cite{She19}; for compactness, in the rest of this discussion we only consider the former metric.

%\textcolor{blue}{
%In order to implement these three main uses, several key performance indicators (KPIs) have been defined by the IUT to measure the characteristics and performance of techniques and technologies proposed: Peak throughput per user; Average speed perceived by the user; Spectral efficiency; Maximum terminal speed; Latency; Number of objects connected to an area; Energy efficiency of the network; Flow over an area; [2a][3a] ; And more, reliability, mobility downtime, bandwidth, maximum spectral efficiency, effective capacity.}\\
%\textcolor{blue}{
\subsection{Which multiple access technique to use?}

\par Regarding the choice of the multiple access (MA) technology itself, faced with the scarcity of resources -- at least until mmWave technologies become commonplace -- %in typical cellular frequency bands, which has led to the exploitation of bands which were hitherto unused, and the upcoming explosion announced of IOT,
the suitable radio MA arises as a stressing issue, particularly in the uplink under massive connectivity. Among the various proposed PHY solutions to increase throughput (and consequently decrease latency), e.g., \cite{Chorti_FTN}, the exploitation of non-orthogonal multiple access (NOMA)  technologies, which until recently have largely remained an information theoretic niche topic, seems the most promising \cite{3GPP_NOMA}.

\begin{figure}[t]
 \centering
 \includegraphics[width=0.5\textwidth]{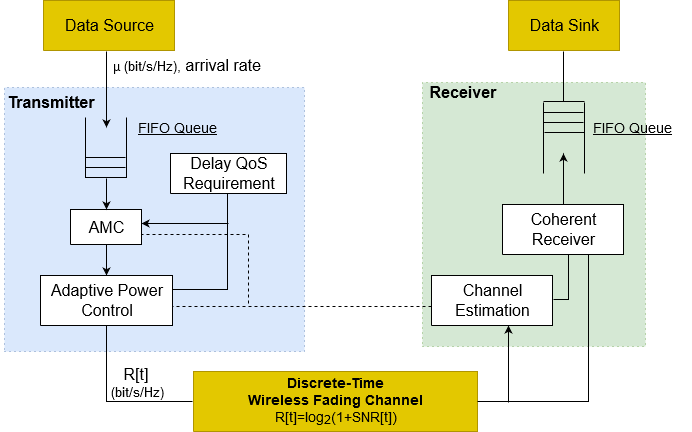}
 \caption{System block diagram}
 \label{queuing}
\end{figure}

\begin{figure*}[t]
   \begin{center}
      \includegraphics[width=0.88\textwidth]{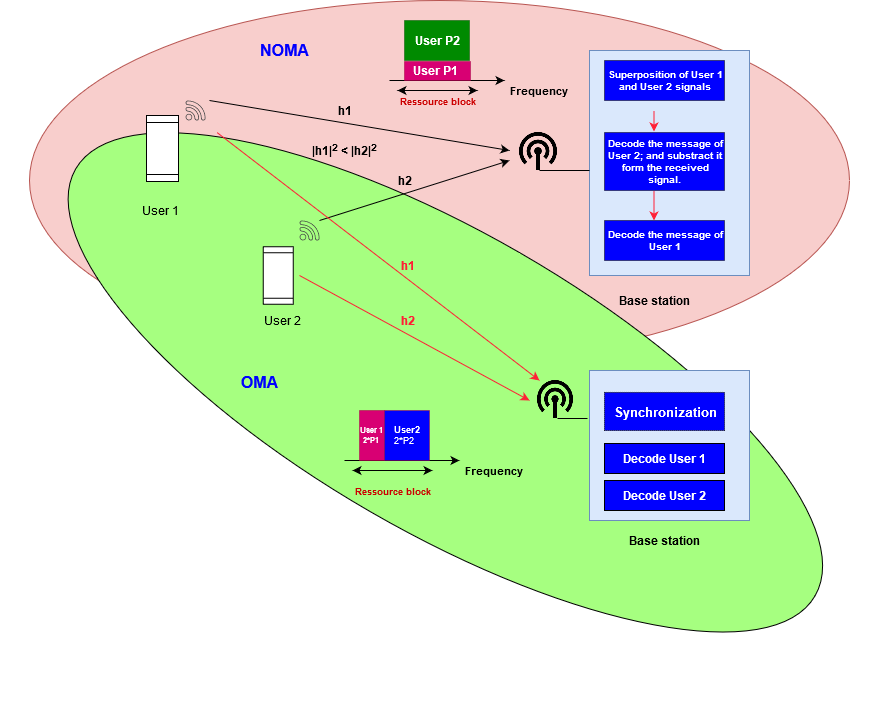}
      \caption{OMA and NOMA MA in a two user network.}
      \label{fig1}
    \end{center}
  \end{figure*}

In power domain NOMA (referred to as NOMA henceforth), several users can use the same resource blocks (RBs) simultaneously by employing superposition coding at the transmitter and successive interference cancellation (SIC) at the receiver, as illustrated in Figure \ref{fig1}. On the contrary, orthogonal MA (OMA) techniques such as time division MA (TDMA) allow only a single user to be served within the same time (or frequency) RB.  NOMA has been shown to outperform OMA and other alternatives in  terms  of  spectral  efficiency, cell-edge throughput, energy efficiency and achievable secrecy rates \cite{SecrecyYuChorti19, Chorti_secFTN}. A natural question that arises is whether the supremacy of NOMA  over OMA holds for the case of delay sensitive applications, with a particular focus on scenarios with users of varying QoS delay requirements.

In this study, we shed light to this question by utilizing the concept of the effective capacity in an uplink network and study OMA / NOMA with or without user pairing. Finally, we take a further step by discussing a novel, flexible MA technique that allows harnessing the benefits of both NOMA and OMA, particularly when users that experience favorable channel conditions require delay sensitive services.

%Discrimination between several users can be done by assigning different codes to each user (Code domain NOMA) or by allowing them to transmit with different powers (Power domain NOMA) in the same resource block (RB).  Until then, orthogonal multiple access techniques: FDMA, TDMA, and OFDMA allow only a single user to be served within the same time/frequency RB; with an exception made by CDMA, which allows multiple users to be supported by the same RB with the aid of applying different unique, user-specific spreading sequences for distinguishing them.%} %\\

%\textcolor{blue}{
%In fact, in B5G (i.e., 6G) QoS following the DiffServ paradigm will be in place everywhere (even for low-end IoT sensors). In particular, in the first 6G white paper it is noted that: `` 6G needs an upgraded networking paradigm moving from best effort to differentiated service quality'' [4]. At the same time, considerable amounts of traffic will be placed on the wireless edge (e.g., ``M2M connections will be more than half of the global connected devices and connections by 2022'', source CISCO [5a]). \\
%.}\\
%\textcolor{blue}{
% NOMA is emerging as a promising technique for future communications and it is increasingly establishing itself as a multiple Acces technique that can meet these challenges or at least some of them.  It is especially important in the case of massive connectivity in B5G.
%This motivated numerous researchers to dedicate substantial research contributions to this new multiple Access technique.}\\

 \section{Effective Capacity in Wireless Networks}

As mentioned above, it is inherently impossible to provide deterministic delay guarantees in radio access due to the random channel variations experienced in the form of small-scale fading and shadowing; intuitively, this is the consequence of the impossibility of a strictly zero outage probability, irrespective of the diversity order employed.

Hence, in this paper, we focus on statistical delay QoS guarantees and employ the theory of EC as a suitable performance metric to confine the delay violation probability below a required threshold. As a dual of the effective bandwidth that was extensively studied for wired networks, the theory of EC was proposed in \cite{wu2003effective}; EC denotes the maximum constant arrival rate that can be supported by a given service process (over wireless channels), on the condition that a target statistical delay QoS requirement is satisfied. Arguably, in most services a variable arrival rate is typical. This issue can be resolved by lower-bounding the EC by the effective bandwidth, as shown in \cite{She19}. In the following, we assume that this requirement holds in all cases without providing further details for the sake of simplicity. %Effective bandwidth was extensively studied in the early 90's with the emphasis on wired networks, which provides the minimum constant service rate that is needed to support a statistical delay requirement for a given arrival process.

To illustrate the concept of EC, as in Figure \ref{queuing}, let us assume a dynamic queuing system with an unconstrained buffer in which the queue length has already converged to a steady state. The probability that the delay experienced by a packet of user $k$ exceeds a maximum delay bound $\Dmax$ has been shown to decay exponentially with the product of three quantities: i) $\Dmax$, ii) the EC, and, iii) the QoS delay exponent, denoted by $\tk$ \cite{wu2003effective}, i.e.,
\begin{align}
   \PoutD&=\Pr\{D_k(t)> \Dmax \} \approx \Pr \{q_k(t)>0\} \mathrm{e}^{-\tk E_c^k \Dmax},
\end{align}
where $ E_c^k$ denotes the EC of user $k$, $\PoutD$ is the delay-outage probability limit for user $k$, $\Dmax$ denotes a maximum tolerable delay and (in units of  symbol periods), $\Pr \{q_k(t)> 0\}$ denotes the probability of a non-empty buffer at time $t$, and $\tk$ is referred to as the delay QoS exponent. % As we can notice from \eqref{eq:Pout},
The delay exponent $\tk\left(\tk>0\right)$ captures how stringent the delay constraint is. A smaller $\tk$ indicates that the user can tolerate a loose delay QoS guarantee (delay tolerant applications), while a larger $\tk$ corresponds to a system with more stringent delay QoS requirements (delay sensitive) \cite{yu2018link,wu2003effective}. %Specifically, when $\tk\rightarrow 0$, it indicates that the $\kth$ user has no delay requirement. When $\tk\rightarrow \infty$, it means that the $\kth$ user cannot tolerate any delay outage, yielding an extremely stringent delay requirement \cite{yu2018link}.

In order to satisfy a target delay-violation probability limit, a source needs to limit its maximum constant arrival rate to  EC \cite{wu2003effective}. %Hence, the EC represents the maximum constant arrival rate that can be supported, while guaranteeing a target value of.
Remarkably, applying the theory of large deviations under the assumption that the G\"{a}rtner-Ellis limit exists, allows us to derive in closed form the expression for the EC of user $k$ in a block-fading additive white Gaussian noise (BF-AWGN) channel, expressed as \cite{yu2018link}
\begin{align}
\ECk &= \dfrac{1}{\beta_k} \ln \left( \mathbb{E}\left[e^{\beta_k R_k}\right] \right) \quad \left(\text{b/s/Hz}\right), 	\label{eq:ECk}
\end{align}
where $\mathbb{E}\left[\cdot\right]$ denotes expectation over the channel gains,  $\beta_k=-\tk T_{\text{f}}B$, denotes the (negative) delay exponent, $T_{\text{f}}$ is the frame duration of the fading-block, $B$ is the bandwidth and $R_k$ is the  achievable rate. From the above discussion, it is apparent that the EC is not only a suitable, but also a flexible metric which can represent various latency requirements.

%\vspace{0.3cm}
\section{NOMA and OMA under Delay Constraints}

In this Section, we discuss the performance of NOMA in low-latency communications, captured by small negative delay exponents. Comparative studies between NOMA and the conventional OMA will reveal that NOMA is more advantageous in terms of the total EC, but is not always superior to OMA in terms of individual ECs. As a result, it is possible that the use of NOMA can penalise a user of a delay sensitive application, a fact that needs to be explicitly accounted for when selecting the MA technique.
\subsection{Two user baseline scenario}

%\begin{itemize}
%    \item NOMA System Model
%\end{itemize}
\par Assume a two-user NOMA uplink network with users U$_1$ and U$_2$ in a Rayleigh BF-AWGN propagation channel, with respective channel gains during a transmission block denoted by  $|h_1|^2<|h_2|^2$, as shown in  Figure \ref{fig1}. In the following, user U$_1$ will be referred to as the weak user and U$_2$ as the strong user. The users transmit corresponding symbols with power $P_1$ and $P_2$ respectively, according to the NOMA uplink principle which imposes $P_1 \leq P_2$ \cite{yang2016uplinknoma}.
%$\mathbb{E}[|S_i|^2]=P_i, i=1,2$.
%Here, $P_i$ is the power coefficient for the user $i$ and normalized transmission powers are assumed
%The received superimposed signal can be expressed as the sum of the contributions of the two users $S_1$ and $S_2$ weighted by their respective channel gains $h_i$ and allocated powers $P_i$, plus an additive zero mean circularly symmetric complex Gaussian noise with variance $\sigma^2$\cite{nzhang2016uplink}, see Fig. XXX. \textcolor{purple}{I suppose it is fine to add one inline equation here, instead of explaining the superimposed signal with a wordy sentence?}
%\begin{equation}
%    Z = \sum_{i=1}^{2}\sqrt{P_i}h_iS_i+w
%\end{equation}
%where \textcolor{black}{$w$} denotes a zero mean circularly symmetric complex Gaussian random variable with variance $\sigma^2$.
The receiver (base station) will first decode the symbol of the strong user treating the signal of the weak user as interference. After decoding it, the receiver will suppress it before decoding the signal of the weak user. Following the SIC principle and denoting by  $\rho$ the transmit signal-to-noise ratio (SNR), the achievable rate, in b/s/Hz, for user
U$_i, i=1,2$, is expressed as \cite{islam2016power}
 \begin{equation}
    R_i= \log_2 \left[1+ \frac{\rho P_i|h_i|^2}{1+\rho \sum_{l=1}^{i-1}P_l|h_l|^2}\right].
    \label{eq_rate}
 \end{equation}

 %\textcolor{red}{Mylene: I removed the fact that power is split between both users as this no longer the case. } \\
 On the other hand, the achievable rate in OMA follows the standard time sharing principle. Furthermore, with respect to the allocation of the RBs, amongst various approaches, when faireness is the prevalent criterion, it is commonplace to assume that U$_1$ and U$_2$ utilize half of the available resources each when using OMA but have access to double the power (same energy expenditure per RB) \cite{Zhinguo16}. As a result, their achievable rates are capped to half their respective Shannon capacities over the totality of the resources assuming twice the power budget. % is denoted $ \tilde{R}_i$. It corresponds to a similar expression as (\ref{eq_rate}) with a denominator equal to $1$ since there is no interference.
 %$ \tilde{R}_i$ is divided by two with respect to $R_i$
%  \begin{equation}
 %   \tilde{R}_i= \frac{1}{2} \log_2 \left[1+ \rho P_i|h_i|^2\right].
 %\end{equation}
 %as each user transmits once every two time transmit interval, and assuming the same maximum transmit power per time transmit interval.

The ECs of either user in either MA approach are evaluated by inserting the respective achievable rates into equation \eqref{eq:ECk}. A detailed illustration of our comparative findings is given below.

\begin{figure}[t]
 \centering
 \includegraphics[width=0.5\textwidth]{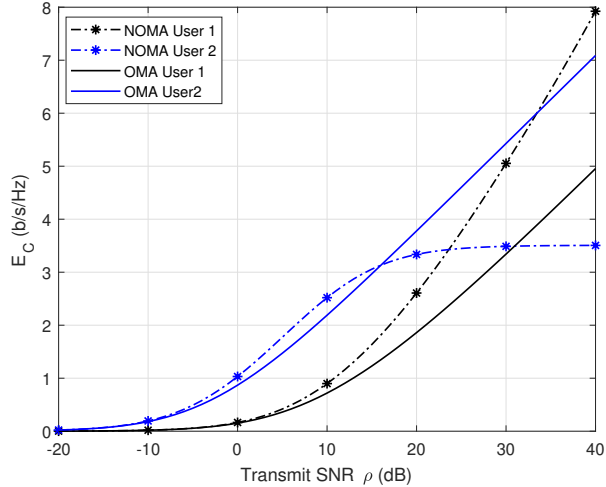}
 \caption{NOMA versus OMA comparison for $\beta$=-5.}
 \label{fig2}
\end{figure}

\begin{figure}[t]
   \begin{center}
      \includegraphics[width=0.5\textwidth]{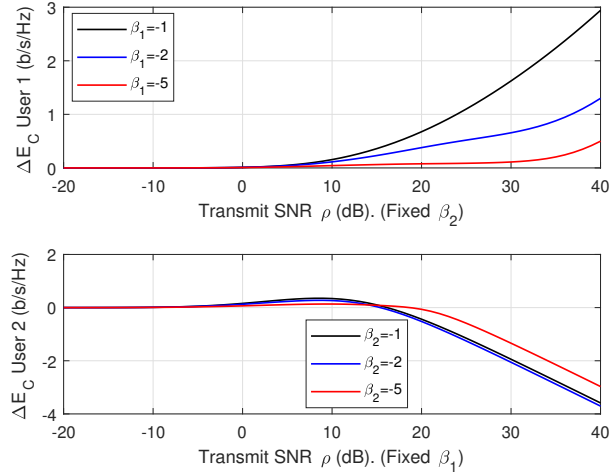}
      \caption{$\Delta$$E_c$ versus the transmit SNR for various negative delay exponents, in a logarithmic scale, for the strong and weak users. }
      \label{fig3}
    \end{center}
  \end{figure}

  \begin{figure}[t]
    \centering
      \includegraphics[width=0.5\textwidth]{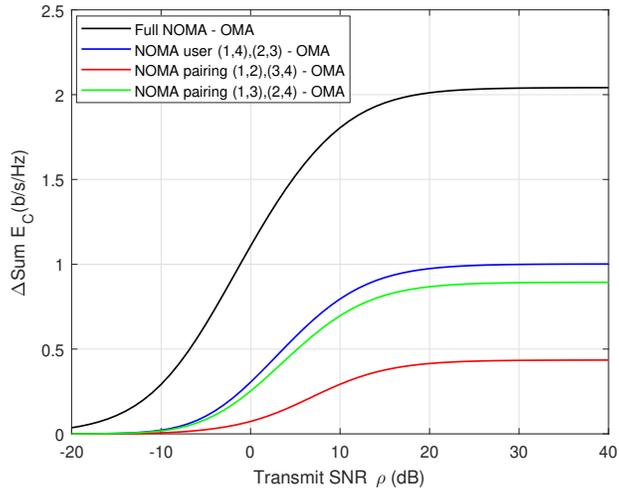}
      \caption{Comparison between OMA, full NOMA and NOMA with user pairs in terms of $\Delta$ Sum $E_{C}$ for $M=4$ and $\beta_1=\beta_2=-5$.}
      \label{fig6}
  \end{figure}

%\begin{figure}[t]
 %   \begin{center}
  %    \includegraphics[width=0.5\textwidth]{Figure5.eps}
%      \caption{Comparison between OMA, NOMA, NOMA user-paired and NOMA grouping 3 users, in term of $\Delta$Sum E$_C$, $M=6$.}
%      \label{fig4}
%      \end{center}
%  \end{figure}

\subsection{Comparison between NOMA and OMA}

\par It is known that NOMA outperforms OMA in terms of the sum spectral efficiency. However, it is not clear how the two MA schemes compare when statistical delay QoS are in place. To address this question, we begin by analyzing the ECs of the two users as a function of the transmit SNR, depicted in Figure \ref{fig2} for a fixed common negative delay exponent $\beta=-5$ for both users and $P_1=0.2$, $P_2=0.8$.
%To investigate the performance of NOMA in supporting delay-sensitive communications, we plot Fig. \ref{fig:CDSA} which include the comparison results between NOMA and OMA in terms of individual EC.
%\textcolor{red}{Mylene: We must revise this section because conclusions may have changed}\\
We  notice that for the strong user, NOMA achieves higher EC than OMA at low SNRs, while OMA is more advantageous than NOMA at high SNRs. On the other hand, for the weak user, NOMA seems to be better than OMA at both low and high transmit SNR. This reveals that NOMA does not always outperform OMA in terms of individual ECs. We also notice that the EC for the strong user reaches a plateau at high SNRs, due to the interference from the weak user, i.e., the MA scheme is interference limited for users with more favourable channel conditions.

Furthermore, to study the performance of NOMA in various delay scenarios, we plot Figure \ref{fig3}. In the upper sub-figure, the difference of the ECs achieved with NOMA and OMA, denoted by $\Delta E_c$, is depicted for U$_1$ and, respectively, in the lower sub-figure for U$_2$, for varying delay exponents. It is shown that for the weak user, as the delay requirement becomes more stringent, NOMA always outperforms OMA. In contrast, the strong user observes a phase transition in medium SNRs; in low SNRs, NOMA outperforms OMA while in high SNRs the opposite is true. As the negative delay exponent decreases, i.e., the delay constraints become more stringent, NOMA outperforms OMA in terms of EC over a larger range of SNRs. However, the gap in the ECs between NOMA and OMA decreases as the latency becomes more stringent. On the other hand, there is a clear penalty in EC at high SNRs when using NOMA; this penalty decreases with decreasing negative delay exponents, i.e., as the delay constraints become stringer.% }

The impact of this initial analysis is worth examining more closely in potential B5G scenarios. As an example, assume a high SNR delay sensitive user\footnote{A high SNR is prerequisite to reach the target low BER for ultra reliability in URLLC, for example.} of a NOMA network, in which it plays the role of the strong user;  in that case, the use of NOMA will result in penalties in terms of the service rate that would be compatible with the statistical delay QoS guarantees. %On the other hand, if the weak user is delay sensitive user is %On the other hand, the URLLC applications are typically low rate,
%\textcolor{purple}{Note that Fig. \ref{fig:CDSA} is plotted with $\beta= ...$, where  $\beta = - \frac{\theta T_f B}{\log(2)}$ is defined as the negative delay exponent. If the absolute value of $\beta$ is large, it corresponds to a user with stringent latency requirements.
%\textcolor{red}{
As a result, the interplay between latency and throughput can be accentuated by the choice of MA.
%\textcolor{blue}{
%The comparison of NOMA with OMA the conventional multiple access technique, on figure 1, - for two user network in the uplink scenario-, shows that NOMA is not always better in terms of effective capacity. That figure shows the individual EC of the weak and strong user. What comes out is that, in low SNRs OMA is better for the weak user and NOMA is better for the strong user. In contrast, in high SNRs NOMA becomes better for the weak user and OMA for the strong one. The Strong user is limited in terms of EC. This is due to interference it experiences from other users weaker than it, as the decoding order is from strongest to weakest at the base station.}\\
As the NOMA objective is to outperform conventional MA methods, it is natural to ask whether it is possible to improve the performance of NOMA compared to OMA in regions where the latter does better than the former when accounting for statistical delay constraints. To address this question, we first explore user-pairing with NOMA, discussed next, and we subsequently introduce a novel scheme, referred to as NOMA-R.

%\begin{figure}[t]
%    \begin{center}
%      \includegraphics[width=0.5\textwidth]{Figure6.eps}
%      \caption{Comparison between OMA, NOMA, NOMA user-pairing and NOMA grouping 3 users, in term of the Sum E$_C$, $M=6$.}
%      \label{fig5}
%      \end{center}
%  \end{figure}

\subsection{Impact of user pairing}

%In Figures \ref{fig4} and \ref{fig5}, the comparison between OMA, NOMA, NOMA user-paired and NOMA with a group of three users (instead of two), is made. In terms of the sum EC, Full NOMA outperforms all other MA approaches, including NOMA with groups of two or three users. Furthermore, all NOMA variants outperform OMA in terms of the sum EC. As a result, in a network of $M$ users, maximizing the sum EC dictates superimposing all users over the RB.

In larger networks with $M>2$, although in terms of the sum EC it is beneficial to superimpose as many users as possible \cite{Zhinguo16}, with respect to the individual ECs this may not be the case. Furthermore, there are practical limitations in the number of users that can be superimposed. Firstly, performing several SICs in series at the receiver may lead to additional processing delay and thus increase the end-to-end (E2E) latency, especially for the weaker users (note that in the uplink the weak users are bound to lower transmit power, which in turn already penalizes their respective rates). Additionally, error propagation can severely impact the system performance in the case of imperfect SIC decoding \cite{Liu19}. %, which may lead to lower rates for the last decoding users.

An alternative MA approach, referred to as user-paired NOMA, has been proposed  due to its advantages in terms of practical implementation. This MA approach is in essence a hybrid of OMA and NOMA; the users are split into a small number of groups that share between them the RBs according to OMA, while within each group the users are superimposed according to NOMA. It is known that pairing does not bring an improvement with respect to sum rates compared to full NOMA. Perhaps one should rather say, that it is a compromise between a variety of conflicting requirements: (a) for weak(er) users to decrease complexity and increase reliability by reducing the SIC layers, at the expense of a reduction in the achievable rate and by consequence in the EC; and (b) for strong(er) users to decrease interference and potentially increase their achievable rates and respective ECs, at the expense of lower sum rates and sum ECs in the network.%\footnote{In multipath rich transmission environments, the roles of weak(er) and strong(er) users can potentially change arbitrarily over a horizon of transmission time intervals due to deep fading, showcasing that the sum EC is a common quantity of interest to all users.}.
%That is to say, we have reasonable delays for weak users by limiting the number of users per RB, and we pay the cost in effective capacity. (which explains the fall of the EC when we decrease the number of users sharing the same resource block).

Studying the impact of user pairing, in Figure \ref{fig6} we depict the gap in the sum ECs (denoted by $\Delta$Sum$E_C$) when using NOMA -- with various pairing approaches -- and OMA, in a $M=4$ user network. Clearly, full NOMA with all users superimposed can in theory outperform all other approaches. However, as argued before, mutli-layer SIC can severely compromise reliability due to error propagation. Alternatively, NOMA with optimal user pairing  provides a reasonable compromise between performance, complexity and feasibility.

For the user-paired NOMA, it is well established that the pairing strategy has a great impact on the system performance. With respect to sum rates, optimal pairing entails grouping the strongest (user 1 in Figure \ref{fig6}) with the weakest user (user 4 in Figure \ref{fig6}), the second strongest (user 2 in Figure \ref{fig6}) with the second weakest (user 3 in Figure \ref{fig6}), and so on \cite{Zhinguo16}. % Figure \ref{fig6} shows the difference of the Sum EC of NOMA user-paired and OMA; and the Full NOMA one; i.e., $SumEC_{\text{NOMA user-paired}} - SumEC_{\text{OMA}}$ and $SumEC_{\text{Full NOMA}} - SumEC_{\text{OMA}}$, for four users.
The same optimal pairing strategy $(1,4),(2,3)$ achieves also a considerable gain in terms of $\Delta$Sum$E_C$, compared to the worst pairing strategy which is sequential (users are paired in decreasing order).

A larger SNR gap is beneficial in terms of the sum ECs, confirming relevant prior results for the case of the achievable rates.
%Maximum benefits in the user pairing are achieved when the strongest user is paired with the weakest one, while the second strongest user is paired with the second weakest one, etc.
Although the user-paired NOMA may achieve a lower sum EC than the full NOMA, it limits the experienced interference at the strong user, as the strong user needs to be decoded first with the weak users' signals are treated as interference. %On the other hand, Figure 6 also reveals that the performance of user-paired NOMA can be further improved by carefully selecting the optimal pairing strategy.

This inspires us to focus on the user-paired NOMA with the optimal pairing strategy.
These results do not account for the individual ECs. We next investigate in further detail this issue and move on to propose a flexible MA strategy which can switch from user-paired NOMA to OMA when the former penalizes the EC of the stronger users.
%\textcolor{blue}{
%From these previous paragraph it's clear that pairing
%shows the difference of EC NOMA user-paired and OMA, $EC_{NOMA} - EC_{OMA}$, for $M=4$ users. We noticed that how to pair users has an impact on the effective Capacity. In fact, in the four users scenario, pairing the strongest user with the weakest one, and the second strongest with the second weakest seems to be the best strategy, i.e the one that maximizes the gab between NOMA EC and OMA EC. That is to say, with pairing you can only do less than pure NOMA in terms of total EC. But it still saves us from the complexity of finding closed-form expressions of the EC, and also it limits the interference on individual EC, faced by the strong users due to weak users; because the earlier ones are decoded first at the base station in the uplink NOMA. We can therefore answer to the question of knowing if by pairing users, we can go beyond pure NOMA, in terms of the total Ec, by the negative. On the other hand, by playing on pairing strategies (power allocation) we think that we can well approach the performance of pure NOMA, as it is shown that a better pairing gives better total EC. However the interrogation still remains for individual ECs. (Can pairing increase individual EC?)} \\

%\textcolor{red}{One or two sentences to make a transition...}

\begin{table*}[t]
\caption{Recommended MA techniques for various scenarios} \label{tb:1}% title of Table
\centering\renewcommand{\arraystretch}{1.4}
\setlength{\tabcolsep}{10.5pt}
\centering
\begin{tabular}{|c|c|c|c|l|}
\hline\hline
\multicolumn{4}{|c|}{Scenarios}                                               & \multirow{3}{*}{Recommended MA techniques} \\ \cline{1-4}
\multicolumn{2}{|c|}{Low latency service} & \multicolumn{2}{c|}{Transmit SNR} &             \\ \cline{1-4}
Weak user   & Strong user    & High  & Low   &     \\ \hline
  \checkmark  &     \checkmark   &      &   \checkmark  & \begin{tabular}{l}
     NOMA-R (optimal pairing)
  \emph{or} NOMA (optimal pairing)
  \end{tabular} \\ \hline
  \checkmark  &     \checkmark   &  \checkmark    &     & \begin{tabular}{l}
     NOMA-R (optimal pairing)
  \emph{or} OMA
  \end{tabular} \\ \hline
  \checkmark  &        &      & \checkmark    &
     \begin{tabular}{l}NOMA-R (optimal pairing)
  \emph{or} NOMA (optimal pairing) \end{tabular} \\ \hline
     \checkmark  &        &  \checkmark    &     &
     \begin{tabular}{l} NOMA (optimal pairing) \end{tabular} \\ \hline
    &    \checkmark    &     &   \checkmark   &
     \begin{tabular}{l} NOMA-R (optimal pairing) \emph{or} NOMA (optimal pairing) \end{tabular} \\ \hline
     &    \checkmark    &  \checkmark    &     &
     \begin{tabular}{l} NOMA-R (optimal pairing) \end{tabular} \\ \hline
       &        &     &  \checkmark   &
     \begin{tabular}{l} NOMA-R (optimal pairing) \emph{or} NOMA (optimal pairing) \end{tabular} \\ \hline
     &        &  \checkmark    &     &
     \begin{tabular}{l} NOMA-R (optimal pairing) \end{tabular} \\ \hline
\end{tabular}
\end{table*}

\section{NOMA-R Strategy for Networks with Strong Low Latency Users}
%Note: I revised the first paragraph to remove conflicting observations. This is because Mouktar and Mylene used different power models. Mylene's observations of "{\it The achievable rate is always larger with NOMA than with OMA for the weak user}" is because of the power model she used, but this has conflicts with Fig. 1 which is plotted by Mouktar who used a different power model. Eventually, we will fix one model but for now, we have to submit the paper using what we have. So I revised the following paragraph to avoid the conflicting observations. (by Wenjuan)

As previously seen, OMA may outperform NOMA in terms of EC for the strong user at high SNRs due to large interference, which can be an issue when the strong user receives low latency services. Furthermore, overall, the achievable rates of strong users can have a greater impact on the total rate performance, as we noticed in \cite{yu2018link}. In order to avoid the interference limitation and to achieve better performance in networks with strong low latency users, a flexible MA strategy, referred to as NOMA-R was proposed in  \cite{PischellaNOMA2019WCL}. With NOMA-R, the  MA technique that maximizes the achievable rate of the strong user is selected flexibly. %More specifically, the selection of MA techniques depends on whether the following criterion is fulfilled:}
%individual achievable rate of both users is selected. The achievable rate is always larger with NOMA than with OMA for the weak user, but it may be lower for the strong user. Therefore the selected MA technique depends on whether the following criterion is fulfilled:}
%\begin{align}
%\label{InitialNOMARelevantCriterion}
%    \left(1+ \frac{\rho P_2 x_2}{1+\rho P_1 x_1} \right) \geq ( 1+\rho P_2 x_2 )^{\frac{1}{2}}
%\end{align}

It can be shown that the probability of choosing NOMA over OMA in NOMA-R, tends to unity in low SNRs  and becomes negligible in high SNRs, in accordance to intuition gained from the initial discussion around Figure 2. The EC in NOMA-R can be obtained from the instantaneous NOMA-R rate, evaluated using a time sharing strategy with weights equal to the corresponding probabilities of choosing NOMA or OMA.
The EC of the strong user is always superior with the NOMA-R strategy compared to both NOMA and OMA, whereas the EC of the weak user with NOMA-R is larger than that with OMA, but lower than that with NOMA, as expected.

\begin{figure}[t]
\centering
 \includegraphics[width=0.45\textwidth]{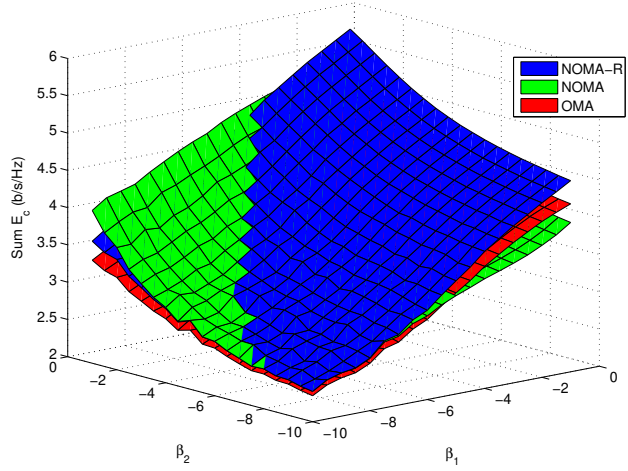}
  \caption{Sum $E_c$ versus $\beta_1$ and $\beta_2$.}\label{3DSumEC_NOMAR}
\end{figure}

\begin{figure}[t]
\centering
 \includegraphics[width=0.45\textwidth]{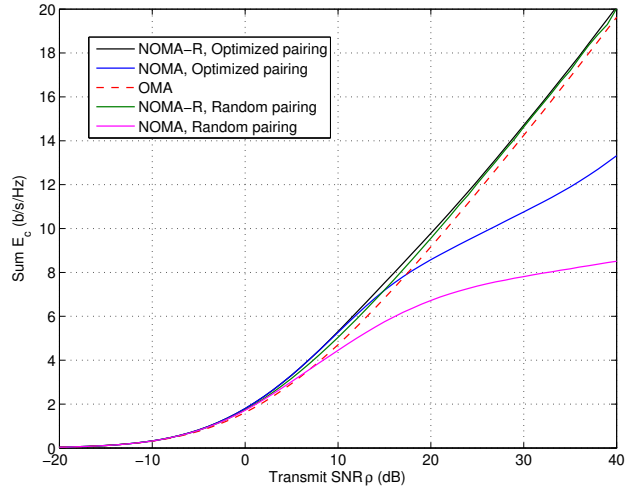}
  \caption{Sum $E_c$  with $4$ users and $2$ pairs, with $\beta_1=\beta_2 = -5$.}\label{SumEC4UsersNOMAR}
\end{figure}

\begin{figure}[t]
\centering
 \includegraphics[width=0.45\textwidth]{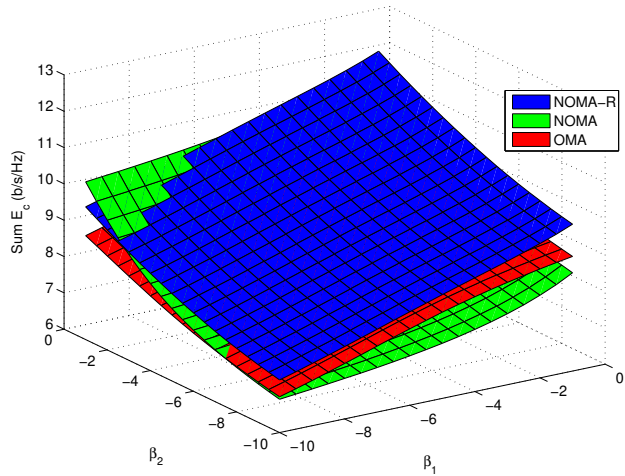}
  \caption{Sum $E_c$ versus $\beta_1$ and $\beta_2$ with $M=4$ users and optimal pairing $(1,4), (2,3)$.}\label{NOMAR4users3D}
\end{figure}
\vspace{0.3cm}
\par \noindent\textit{\textbf{NOMA-R in a two user baseline network}}\\
\vspace{-0.2cm}
\par The influence of the delay QoS exponent parameters $\beta_1$ and $\beta_2$   is represented in Figure \ref{3DSumEC_NOMAR}, where the transmit SNR is set to $20$ dB and $P_1= 0.2$, $P_2=0.8$. NOMA-R is well-suited for stringent target delay-bound violation probabilities, that is, when $\beta_1$ and $\beta_2$ decrease. Adapting the MA strategy in order to avoid interference-limited situations for the strong user allows to better fulfill delay constraints. The proposed NOMA-R flexible MA selection strategy may consequently be a strong candidate for URLLC applications.

\subsection{NOMA-R in a multi-user network}

\par Moreover, users pairing can be optimized according to the NOMA-R criterion. Users are paired in order to maximize the the signal to interference and noise ratio of the strongest user.  This heuristic MA strategy aims at increasing the probability to select NOMA with NOMA-R for the users that are likely to have the largest ECs.

The obtained sum EC with NOMA-R, NOMA and OMA with either random or optimized pairing when $P_1= 0.2$, $P_2=0.8$ and $\beta_1 = \beta_2 = -5$ is represented in Figure \ref{SumEC4UsersNOMAR}.  NOMA-R is the best technique with optimized pairing at large SNRs, while NOMA is shown to be particularly inefficient with random pairing due to large interference at the strong user. Combining NOMA-R with optimized pairing therefore provides the largest sum EC.

Finally, the influence of $\beta_1$ and $\beta_2$ on the EC with the same parameters as before is depicted in Figure \ref{NOMAR4users3D}. A comparative analysis of Figures \ref{NOMAR4users3D} and \ref{3DSumEC_NOMAR}  shows that the $\mathbf{\beta}$ region where NOMA-R outperforms NOMA in terms of sum EC increases thanks to pairing. Please notice that the z axis on Figures \ref{NOMAR4users3D} and \ref{3DSumEC_NOMAR} differs and that the achieved data rate with $4$ users pairing is more than twice that with $2$ users. The sum EC is lower with NOMA-R
than with NOMA with increasing $\beta_2$, i.e., as the strong user tends to receive a delay tolerant service.

The recommended MA techniques for various scenarios (depending on the existence of low latency service for each user and the value of transmit SNR) are summarized in Table \ref{tb:1}. Different latency scenarios are discussed: 1) when both users require low latency services; 2) when only one user has low latency requirements; 3) when both users are delay tolerant. We can conclude that when both users require low latency services, NOMA-R serves as the best MA technique, no matter whether the transmit SNR is high or low.

%\textcolor{blue}{Wenjuan: Table I includes the recomended MA techniques for various delay and SNR scenarios. For the second row, i.e., weak user low latency, strong user, low latency, the conclusion is obtained by considering Fig. 7 and Fig. 10. For the third row, the conclusion is given by considering Fig. 10. Since $\rho=20$ dB in Fig. 10, so this means only the case of high SNRs are discussed. For the fourth row, the conclusion is obtained by considering Fig. 10.}

   \section{Conclusion}

 In B5G, delay constraints emerge as a topic of particular interest, e.g., for uHSLLC and URLLC services such as autonomous vehicles, enhanced reality, factory automation, etc,. In this context, novel MA techniques from the realm of NOMA have attracted a lot of attention in recent years. In this contribution, we provide a comparative performance analysis for the uplink of networks with different MA strategies, when statistical delay QoS  constraints are in place. The latter are captured through each user's effective capacity, a MAC layer rate metric that accounts for QoS delay exponents.

 The first outcome of our analysis shows that in the high SNR region, "strong" NOMA users are interference limited, which translates in plateaus in their ECs. We further show that the inverse conclusions hold for the weak users.
 The second conclusion reached concerns the impact of user pairing or grouping and its impact on the ECs. Due to practical considerations, particularly in view of imperfect SIC, it is indicated that NOMA with optimal user pairing is a promising implementation approach for NOMA as it provides a good compromise between performance and feasibility.

 %Furthermore, given the ultra reliability constraints for URLLC services, a high SNR is naturally required.
 Finally, in networks with users of varying delay QoS requirements, it is probable that the strong users are recipients of low latency services, e.g., they are URLLC users. In this scenarios we propose a flexible MA strategy, referred to as NOMA-R, in which NOMA is only adopted when it benefits the strong users. Our numerical results show that NOMA-R offers concrete benefits even in terms of sum ECs, particularly when the strong user has stringent delay QoS constraints.

 %\section*{Acknowledgment}
%This work was supported in part by the INEX project eNiGMA and the ANR project ELIOT.

 %\section*{REFERENCES}\\
 %\textcolor{blue}{
 %1a- \url{https://www.itu.int/en/mediacentre/backgrounders/Pages/5G-fifth-generation-of-mobile-technologies.aspx}\\
 %2a- \url{https://5g-ppp.eu/wp-content/uploads/2016/11/06_10-Nov_Session-3_Lee-JunHwan.pdf}\\
 %3a- \url{https://www.itu.int/dms_pubrec/itu-r/rec/m/R-REC-M.2083-0-201509-I!!PDF-E.pdf}\\
 %4a- \url{https://www.oulu.fi/6gflagship/news/6g-white-paper.} \\
% 5a- \url{https://www.cisco.com/c/en/us/solutions/collateral/service-provider/visual-networking-index-vni/white-paper-c11-741490.html}
% }

%\ifCLASSOPTIONcaptionsoff
%  \newpage
%\fi
% \vspace{-2mm}
 \bibliographystyle{plain}
 \bibliography{HuaEE}

 % add ref https://www.oulu.fi/6gflagship/news/6g-white-paper

 %%%%%%%%%%%%%%%%%%%%%%%%%%%%%%%%%%%%%%%%%%%%%
 % we don't need to provide bios now. Those can be added after it is accepted.
 %\newpage
%\begin{IEEEbiography}[{\includegraphics[width=1in,height=1.25in,clip,keepaspectratio]{Author.jpg}}]{Mouktar~Bello} (S'09-M'09)
%\end{IEEEbiography}
%\begin{IEEEbiography}[{\includegraphics[width=1in,height=1.25in,clip,keepaspectratio]{Author.jpg}}]{Wenjuan~Yu}  (S'11, M'13)
%\end{IEEEbiography}
%\begin{IEEEbiography}[{\includegraphics[width=1in,height=1.25in]{Author.jpg}}]{x~y}  (S'15 - M'16)
%\end{IEEEbiography}
%\begin{IEEEbiography}[{\includegraphics[width=1in,height=1.25in,clip,keepaspectratio]{Author.jpg}}]{x~y} (M'03-SM'12) \end{IEEEbiography}
%\begin{IEEEbiography}[{\includegraphics[width=1in,height=1.25in,clip,keepaspectratio]{Author.jpg}}]{Inbar~Fijalkow}
%\end{IEEEbiography}

% that's all folks
\end{document}